\documentclass[11pt,twoside]{article}  
\usepackage{apn3conf}

\begin{document}   

%
%
%
%

\title{NLTE Spectral Analysis of Central Stars of PNe Interacting with the ISM}

%
%
%

\author{Thomas Rauch}
\affil{Dr.-Remeis-Sternwarte, Bamberg, Universit\"at Erlangen-N\"urnberg, Germany}
\affil{Institut f\"ur Astronomie und Astrophysik, Universit\"at T\"ubingen, Germany}
\author{Florian Kerber}
\affil{Space Telescope - European Coordinating Facility, Garching, Germany}
\author{Elise Furlan}
\affil{Center for Radiophysics and Space Research, Cornell University, Ithaca, NY, USA}
\author{Klaus Werner}
\affil{Institut f\"ur Astronomie und Astrophysik, Universit\"at T\"ubingen, Germany}
%
%

\contact{Thomas Rauch}
\email{Thomas.Rauch@sternwarte.uni-erlangen.de}

%
%
%
%
%

\paindex{Rauch, T.}
\aindex{Kerber, F.}     
\aindex{Furlan, E.}     
\aindex{Werner, K.}     

%
%

\authormark{Rauch et al.}

%
%



\begin{abstract}          
The analysis of Planetary Nebulae (PNe) provides
a tool to investigate the properties of their exciting central
stars (CSPN) at the moment of the PN ejection as well as on the properties
of the ambient interstellar medium (ISM). The spectral analysis of the CSPN
is a prerequisite to calculate the ionizing flux which is a crucial input
for reliable PN modeling.
In the framework of a systematic study of PNe interacting with the ISM (Kerber et al\@. 2000), 
we present preliminary results of ongoing NLTE spectral analyses of ten 
of their CS based on new optical medium-resolution spectra.
\end{abstract}

%
%

Recent developments in observational and deprojection techniques, spectral
analysis, and numerical methods facilitate to closely examine and model
PNe and their CS. These stars are at their
hottest stage of evolution close to the end of nuclear burning, and
gravitational effects become dominant, i.e\@. they display directly the
formation of white dwarfs. 

An indicator for their evolution is 
the interaction of the associated PN with the ambient ISM: 
the highly evolved CS is no longer dominating the processes in the PN
(Kerber \& Rauch 2001); the nebula displays an asymmetric brightness distribution that 
reflects the degree of the interaction process.
These complex objects are crucial tests for our models as well as evolutionary
theory.

Spectral analysis of CSPN by means of NLTE model atmosphere techniques
provides information about photospheric parameters like $T_\mathrm{eff}$,
$\log g$, and surface abundances.
In comparison with evolutionary calculations, we can then determine the
evolutionary status, distance, mass, and luminosity of the CSPN. Moreover, the model fluxes
can be used as realistic ionizing spectra in analyses of the PNe. 
These allow us to obtain detailed information on the ionization structure 
of the nebulae, particularly in their complex interaction zones.

\begin{table}[ht]
\caption{Parameters of our programme stars. Detailed analyses
of the CSPN of DeHt\,5 and EGB\,1 have been presented by 
Barstow et al\@. 
(2003, $T_\mathrm{eff}\hspace{-0.5mm} =\hspace{-0.5mm} 58\,582\,\mathrm{K}$ and $\log g\hspace{-0.5mm} =\hspace{-0.5mm} 7.05$) 
and Napiwotzki 
(1999, $T_\mathrm{eff}\hspace{-0.5mm} =\hspace{-0.5mm}    147\,\mathrm{kK}$ and $\log g\hspace{-0.5mm} =\hspace{-0.5mm} 7.34$),
respectively.
The CSPN of A\,21 and RX\,J2117.1+3412 have been analyzed by Rauch \& Werner (1995) and 
Rauch \& Werner (1997), respectively.
Stanghellini et al\@. (2002) presented Zanstra temperatures for the CSPN of 
NGC\,6842 ($T_\mathrm{eff}\hspace{-0.5mm} =\hspace{-0.5mm}  97\,\mathrm{kK}$),
A\,75     ($T_\mathrm{eff}\hspace{-0.5mm} <\hspace{-0.5mm} 290\,\mathrm{kK}$),
and
NGC\,6781 ($T_\mathrm{eff}\hspace{-0.5mm} =\hspace{-0.5mm} 105\,\mathrm{kK}$).
}\vspace{2mm}
\label{php}
\begin{tabular}{llccc}
\hline
\noalign{\smallskip}
name & PNG & $T_\mathrm{eff}$ / kK & $\log g$ (cgs) & H/He (mass)\\
\noalign{\smallskip}
\hline
\noalign{\smallskip}
DeHt\,5          & 111.0$-$11.6 &  70 & 7.0 & $>$100 \\
EGB\,1           & 124.0+10.4   & 120 & 8.0 & $>$100 \\
\hline
\noalign{\smallskip}
NGC\,6842        & 065.9+00.5   &  80 & 5.0 &   4    \\
A\,75            & 101.8+08.7   &  80 & 6.0 &   0.7  \\
NGC\,6781        & 041.8$-$02.9 &  80 & 6.0 &   0.7  \\
WeSb\,5          & 058.6$-$05.5 &  80 & 6.0 & $<4$   \\
Sn\,1            & 013.3+32.7   & 100 & 5.0 &   0.4  \\
A\,52            & 050.4+05.2   & 110 & 6.0 &   0.25 \\
\hline
\noalign{\smallskip}
A\,21            & 205.1+14.2   & 140 & 7.5 & \multicolumn{1}{l}{He:C:O=35:51:14} \\
RX\,J2117.1+3412 & 080.3$-$10.4 & 180 & 6.1 & \multicolumn{1}{l}{He:C:O=38:56:6}  \\
\hline
\end{tabular}
\end{table}

In July 1999, we performed medium-resolution spectroscopy of nine CSPN with
the TWIN spectrograph attached to the 3.5m telescope at Calar Alto, Spain.
The CS of A\,21 was observed in January 1999 with EFOSC\,1 at the 3.6m
telescope of ESO (La Silla). Data reduction was carried out using IRAF.
The observed spectra have a resolution of 2.7\AA\ and S/N ratios from 10 to 30. 
Since the nebular emission is highly asymmetric across the projected face of these PNe, the emission cannot be subtracted
perfectly. Especially in the case of Sn\,1, there exists a small (about 8'' diameter), compact inner nebula
which makes a proper background subtraction impossible. However, the line wings can still be used for $\log g$
determination, and He\,{\sc ii}~$\lambda 4199.8\,$\AA\ and He\,{\sc ii}~$\lambda 4541.6\,$\AA\ 
yield information about the H/He ratio. However, the
determination of $T_\mathrm{eff}$ is uncertain in such a case.

The spectral analysis is performed by means of NLTE model atmosphere techniques
employing our code {\tt PRO2} (Werner 1986, 1988, Werner \& Dreiz\-ler 1999).
The models are plane-parallel and in hydrostatic and radiative equilibrium.
In general, {\tt PRO2} is able to treat more than 300 individual atomic levels in NLTE
with more than 1\,000 individual lines (Rauch 1997, 2003)

For the classification and preliminary analysis of hot compact stars, we have set up a
new grid of H+He NLTE model atmospheres with $T_\mathrm{eff}\hspace{-0.5mm} =\hspace{-0.5mm} 50 - 190\,\mathrm{kK}$ 
(in 10\,kK steps), $\log g\hspace{-0.5mm} =\hspace{-0.5mm}  5 - 9$ 
(in 0.5 steps) in cgs units, and H/He from pure H to pure He. With this new grid, we aim to
minimize the error to about 20\,kK in $T_\mathrm{eff}$, 0.5\,dex in $\log g$, and 0.5\,dex
in H/He. The preliminary results of our analysis are summarized in Table~\ref{php}.
The grid which we used for this analysis as well as some other grids of NLTE model atmosphere fluxes with different chemical
composition will be available at
{\tt http://astro.uni-tuebingen.de/\raisebox{1mm}{$\sim$}rauch}.

Within our sample of ten CSPN, whose nebulae show interaction with the ISM, our preliminary classification and
spectral analysis yield 
two hydrogen-rich DA \mbox{(pre-)} white dwarfs (DeHt\,5 and EGB\,1 -- both already known, Napiwotzki 1999), 
two hydrogen-deficient PG\,1159 stars (A\,21 and RX\,J2117.1+3412 -- both already known, Rauch \& Werner 1995), 
and six CS with intermediate H/He ratios (from 0.25 to 4 by mass). 

Fine tuning of the parameters in the next part of this analysis will enable us to determine e.g\@. their spectroscopic
distance reliably (cf\@. Napiwotzki 2001, DeHt\,5: $d = 510 \mathrm{pc}$, EGB\,1: $d = 650 \mathrm{pc}$).  
However, the analysis is hampered by the relatively poor seeing (1\farcs 6 - 2\farcs 5) during the observations. 
Further high S/N optical and UV observations with better spatial resolution would significantly reduce
the error ranges. 

\paragraph{Acknowledgements.}
This research was supported by the DLR under grants 50\,OR\,9705\,5 and 50\,OR\,0201,
and by the DFG under grants RA\,733/3-1 and RA\,733/14-1.

%
%
%
%


\end{document}